\documentstyle[12pt]{article}
\begin{document}

\begin{center}
{\bf ON SUPERSELECTION RULES FOR MACROSCOPIC OBJECTS}
\end{center}

\begin{center}
{\bf L.V. Prokhorov} \\
{\small \it V.A. Fock Institute of Physics, Sankt-Petersburg State
University, 198504 Russia}\end{center}

E-mail: lev.prokhorov@pobox.spbu.ru

\noindent
It is shown that for "ideal" macroscopic objects there are superselection
rules forbidding superpositions of macroscopically distinguishable states of
the objects. For real macroscopic bodies the notion of "weak"
superselection rules is introduced. Some other aspects of the measurement
problem are discussed. \\


\vspace{0.5cm}
\noindent
{\bf 1.}
The problem of measurement in quantum mechanics is connected with
a number of "sub-problems" which are rather difficult by themselves. At the
beginning of the quantum era the problem of description of the
measurement process was connected with the idea that the measuring apparatus
is a classical object while the objects of measurement are quantum ones. So,
which mechanics should be used --- quantum or classical?

Thus, the main problems at 1920' - 1930' were connected with measuring
apparatuses, objects of measurement and quantum mechanics by itself:\\
1) Macroscopic bodies are complex objects from point of view of quantum
mechanics.\\
2) Microsystems are complex objects from point of view of
classical physics.\\
3) The notion of probability amplitude (wave function) was not well
understood both in physics and mathematics.  [Indeed, it seemed at that time
that there was no physical reality behind it (according to [1] "...  every
element of the physical reality must have a counterpart in the physical
theory."); on the other hand, there is no mathematical theory of stochastic
processes using probability amplitudes.]

Now it is clear that all the objects in the Universe are described by
quantum mechanics (QM) (because all of them are some excitations of quantum
fields). Thus, the process of measurement should be described quantum
mechanically. But it does not makes the task much easier mainly because of
wave function---probability amplitude problem in QM. Gradually it became
clear that notions of wave function and particle cannot be understood
without quantum field theory (QFT): evidently, particles are one-particle
excitations of fields, and wave functions characterize these excitations
[2]. It means that "particles" (electrons, photons etc.) are not pointlike
objects --- they are non-local excitations of fields; {\it it is the fields
interactions that are local}, indirectly suggesting the idea that particles
are "material points". Wave functions describe the corresponding excitations
of the fields, i.e. fields are "elements the physical reality", while wave
functions are their "counterparts in the physical theory" (in QM) [2,3]. As
for the probability amplitudes, their nature was elucidated in [4]. But
description of measurement is a more subtle issue.

The process of measurement in a way  is analogous to the scattering of a
microscopic object $o$ by the measuring apparatus $A$:
\begin{equation}
|o\rangle|A\rangle\rightarrow\sum_k c_k|o_k\rangle|A_k\rangle.
\end{equation}
It should be described by quantum mechanics, so $|o\rangle$, $|A\rangle$
are initial and $\sum_k c_k|o_k\rangle|A_k\rangle$ --- final state vectors of
the object and the apparatus ($c_k$ --- complex numbers). The final state of
measuring apparatus and the microscopic object (the sum in (1)) is called an
entangled state. But $A$ serves as a measuring instrument if it can be
found only in one of the states $|A_k\rangle$ of the sum --- then an
experimenter concludes that the microsystem is in the state $|o_k\rangle$.
As is easily seen, the r.h.s. of (1) cannot be written as
$|o^\prime\rangle|A^\prime\rangle$ (otherwise the result of the experiment
would be unambiguous). The act of measurement can be successful only if
the final state is not pure, i.e. if it is described by $\rho$-matrix

\begin{equation}
\hat\rho=\sum_k |c_k|^2 |o_k\rangle|A_k\rangle
\langle A_k| \langle o_k|.
\end{equation}
It means that the state $|A_k\rangle$ appears with probability $|c_k|^2$.

In the process of measurement there are two stages: measurement as a
physical process (the transition (1), pure quantum description), and taking
data by an experimenter --- registration of positions of a pointer (pure
classical procedure). The latter can be done only if the apparatus (pointer)
is in a certain state, say, $k$, i.e. if it is not in the entangled state
(1).  The peculiarity of situation is in the controversy: according to QM
the final state in (1) is an entangled one, while according to the routine
practice  the apparatus $A$ with some probabilities $|c_k|^2$ is always only
in one of the states $|A_k\rangle$. It means that the apparatus is in the
mixed state described by $\rho$-matrix (2).

Transformation of a pure state into the mixed one is called decoherence.
There are two simple examples of such transformation.

1. {\it Influence of environment}. The macroscopic apparatus is a complex
object and its interaction with the outer world cannot be negligible
(because of smallness of intervals between its energy levels) [5]. So, one
has to average over the states of environment.

2. {\it Infrared radiation}. Any body with non-zero temperature radiates
unregistered infrared photons and gravitons (as a result of collisions of
particles composing the body). Averaging over infrared quanta also
transforms a pure state into a mixture.

The transition from the pure state (1) to the mixed state cannot be the
result of intervention of the experimenter (e.g. at the final stage of
taking data). The idea that it is the experimenter who is responsible for
the non-linear operation $\psi\rightarrow |\psi|^2$ was considered long ago
[6] ($\psi$ is a probability amplitude or wave function). But it cannot be
considered as a satisfactory solution of the problem, because e.g., (i)
estimation of star radiation presumes knowledge of some cross sections, i.e.
transition $\psi\rightarrow |\psi|^2$ should take place in stars, but there
are no observers there; and mainly because (ii) quantum mechanics should
(and can) be formulated in such a way that notions of probability amplitudes
and probabilities enter into the theory from the very beginning [2,7], and
one has no need in special agents for passing from amplitudes to
probabilities.

The transition (1) $\rightarrow$ (2) cannot be due to the environment or the
infrared radiation too; both of them introduce decoherence, but they cannot
solve the problem of measurement because they cannot introduce the mixture
of type (2) allowing to fix the state of apparatus $|A_k\rangle$. They would
rather be appropriate for measuring the states of outer world or of
unregistered radiation (see also [8]). Furthermore, the act of measurement
should be meaningful by itself, irrespective to environment or anything else
--- just as a sign of self-consistency of quantum mechanics.

It may seem that appearance of mixture (2) is an example of standard
transition $\psi\rightarrow |\psi|^2$. Actually, there are two different
aspects here. Indeed, if the sum (1) consists of a single term then one
has only the problem of transition $\psi\rightarrow |\psi|^2$; this is not a
problem at all. The real problem arises when there is more than one term in
(1). Then the apparatus must be only in one state, say $k$, with probability
$|c_k|^2$ (mixed state). How can it be that the entangled state (1)
becomes the mixture given by (2)? This is possible if there are
superselection rules (SR) for state vectors $|A_k\rangle$. In this case
superposition of vectors $|A_k\rangle$ with different $k$ is forbidden (or,
at least, it cannot appear in the process of measurement), and the final
state in (1) should be a mixture.  Thus, this problem of measurement reduces
to the proof that for macroscopically distinguishable states of macroscopic
objects there exist superselection rules.

We shall show that for macroscopic objects there are SR
prohibiting superpositions of macroscopically distinguishable states of the
objects (e.g. states with different centers of mass). It solves the problem
of macroscopic apparatuses decoherence in measurement. But first we discuss
the issue of superselection.

{\bf 2.} Mathematically,
the phenomenon of superselection can be defined as a restriction:
superpositions of some state vectors are forbidden, i.e. there are no such
vectors in the formalism, they cannot appear in the process of evolution of
the system. There are two well known examples of SR. \\
1) The vector $\psi=\psi_b+\psi_f$, where $\psi_b$, $\psi_f$ are
correspondingly bosonic and fermionic state vectors, is forbidden. Under
$2\pi$-rotation of the coordinate system the wave function $\psi_f$ changes
the sign ($\psi_f\rightarrow -\psi_f$), and
$\psi\rightarrow\psi^{\prime}=\psi_b-\psi_f$. But the $2\pi$-rotation is an
identical transformation and $\psi$ cannot be changed, i.e. the vector $\psi$
cannot be realized [9]. \\
2) Electric charge gives another example of SR: superpositions of states
with different electric charges are forbidden. It follows, in fact, from
gauge invariance of QED [10]. The latter case is connected with QFT. Notice
also that a physical operator cannot transform  $\psi$ into $\psi^*$, so
there are SR for these states ($c_1\psi+c_2\psi^*$). As for other examples
of SR, see [11].

In QM, if vectors $\psi_1,\psi_2$ belong to a Hilbert space ${\cal H}$, then
$\psi=c_1\psi_1+c_2\psi_2\in{\cal H}$. There can appear the interference
terms $(\psi_1,\psi_2)$, there exist physical operators $\hat A(q,p)$
transforming $\psi_2\rightarrow\psi_1$, i.e. $(\psi_1,\hat A(q,p)\psi_2)\neq
0$.

In QFT there are unitary non-equivalent representations of canonical
commutation relations. Any physical theory is formulated only in one of
these (separable) Hilbert spaces defined by their ground states (cyclic
vectors). There is no physical reality behind the superpositions of vectors
from different separable Hilbert spaces ${\cal H}^{(i)}$, so, there are SR.
The "empirical" rule is: if there is a non-trivial operator $\hat{\cal
S}$ commuting with all the physical operators of a system (Hermitean
polynomials of canonical variables), then the linear combinations of
eigenfunctions of $\hat{\cal S}$ with different eigenvalues cannot be
realized. The eigenvalues of $\hat{\cal S}$ distinguish the spaces ${\cal
H}^{(i)}$. There is no physical operator $\hat A$ in ${\cal H}^{(1)}$,
such that $\hat A\psi^{(1)}=\psi^{(2)}$, $\psi^{(i)}\in {\cal H}^{(i)}$.
Only operators $\hat {\cal A}$, $[\hat {\cal A},\hat{\cal S}]\neq 0$ can
transform the spaces one into another.

Ferromagnetics give the simplest example of the corresponding Hilbert
spaces. Let functions $\Psi^{(i)}=\prod_1^N \psi_n^{(i)},\ \ i=1,2$,
describe a ferromagnetic sample with different directions of magnetization
($\psi_n^{(i)}$ are the spin wave functions of electrons).  State vectors
$\tilde\Psi^{(i)}=\prod_1^m\tilde \psi_k^{(i)}\prod_{m+1}^N \psi_r^{(i)}$
($m$ may be arbitrary large but finite when $N\rightarrow \infty$) describe
excitations of the "vacuum" states $\Psi^{(i)}$, $\tilde \psi_k^{(i)}
\neq \psi_k^{(i)}$. The corresponding Hilbert spaces ${\cal H}^{(i)}$ are
orthogonal because $|(\psi_n^{(1)},\psi_n^{(2)})|=\eta < 1$ and
\begin{equation}
(\tilde\Psi^{(1)},\tilde\Psi^{(2)})= \prod_1^m(\tilde \psi_k^{(1)}, \tilde
\psi_k^{(2)}) \prod_{m+1}^N (\psi_r^{(1)}, \psi_r^{(2)})\sim c_m\eta^{N-m}
\rightarrow 0,\ \ N\rightarrow \infty .
\end{equation}
Vectors $\tilde \Psi^{(1)}, \tilde \Psi^{(2)}$ belong to the orthogonal
Hilbert spaces, and superpositions of these state vectors has no sense, i.e.
we come to SR. Vectors $\tilde \Psi^{(1)}$ cannot be transformed into
vectors $\tilde \Psi^{(2)}$ and vice versa by any physical operator in
${\cal H}^{(1)}$ or ${\cal H}^{(2)}$.  These two states of a ferromagnetic
are macroscopically distinguishable and in principle this sample can serve
as a measuring apparatus; the direction of magnetization plays the role of a
pointer. Analogous statement is valid for other macroscopic objects.

In relativistic QFT (in the Fock space) superpositions of states with
different numbers of particles are physical by definition. These
multiparticle spaces are subspaces of a bigger separable Hilbert space.
Creation and annihilation operators transform these states one into the
other.

{\bf 3.}
The superpositions of macroscopically different quantum states of
macroscopic objects are senseless. We give a formal proof of this statement
for a body with different centers of mass.

{\it Definition.} The macroscopic object is that having all the properties
of a compact stable system of $N$ particles retained in the limit
$N\rightarrow \infty$.

{\it Remark.} Physicists always operate only with finite numbers of degrees
of freedom. Even in QFT they use separable subspace (the Fock space) of
non-separable Hilbert space (the von Neumann space). According to the
Weierstrass theorem any function can be approximated by polynomials. Thus,
any physical operator can be presented by a polynomial of canonical
variables.

THEOREM. Superpositions of the wave functions of a macroscopic object with
different centers of mass are forbidden.

PROOF. Let ${\bf x}_i, {\bf p}_i$, $i=1,2,...,N$, be the canonical
variables of a compact stable system of $N$ particles in the 3D space.
Then
\begin{equation}
\hat {\bf X} =\frac{1}{N}\sum_{i=1}^{N}\hat {\bf x}_i
\end{equation}
is the operator of
center of mass, and the limit $N\rightarrow \infty$ obviously exists. In this
limit $\hat {\bf X}$ commutes with all physical operators $\hat P_n
(\hat{\bf x}_i, \hat{\bf p}_i)$ (polynomials, $n<\infty$)
\begin{equation}
[\hat {\bf X},\hat P_n]\rightarrow 0,\ \ N\rightarrow \infty .
\end{equation}
For simplicity we prove the statement for 1D space. The physical operators
are connected with arbitrary large but finite number $m$ of operators $\hat
x_{i_r}, \hat p_{i_r},\ \ 1\leq r\leq m < N$, e.g. in the series
\begin{equation} \hat P_n (\hat x_i, \hat
p_i)=\sum_{r=1}^{m}\sum_{k=0}^{n}C_{rk}(\hat x_{i})\hat p_{i_r}^k
\end{equation}
there is only $mn+1$ members (the first terms in (6) are linear in $\hat
x_{i_r}, \hat p_{i_r}$;  $C_{rk}(\hat x_{i})$ are polynomials,
$C_{r0}(\hat x_{i})=C_0(\hat x_{i})$). In the commutator
\begin{equation}
\hat Q_{n-1}=\sum_{i=1}^N [\hat x_i,\hat P_n]
\end{equation}
there is only $mn$ additives, so $\hat Q_{n-1}/N\rightarrow 0$, $N
\rightarrow \infty$ (true for any matrix element of $\hat Q_{n-1}$).  This
consideration can be easily extended to the case of 3D space. Thus, $\hat
{\bf X}$ is an operator of the type $\hat{\cal S}$, and superposition of the
$\hat {\bf X}$ eigenvectors with different eigenvalues is forbidden.
Evidently, analogous statement can be proved for states of macrosystems
obtained one from another by rotation (instead of translation). That is what
says mathematics.

The physics behind the phenomenon of decoherence is connected with the issue
of wave function. Indeed, let $\hat a(f)^+$ be the creation operator of some
free scalar field: $\hat a(f)^+|0\rangle=\hat\varphi(f)|0\rangle=|f\rangle$,
$\hat \varphi(f)=-i\int d^3x(f(x) \partial_0 \hat \varphi(x)-\partial_0 f(x)
\hat \varphi(x))$, $f$ is by definition the wave function of the particle
in the state $|f\rangle$.  Suppose that $f=f_1+f_2$, and
$f_1f_2=0$. The functions $f_1,f_2$ are non-zero in domains
$\Omega_1,\Omega_2$, which do not intersect.  Mathematics admits it if
$||f||^2=||f_1||^2+||f_2||^2=1$. Physically it means that the field is
excited only in domains $\Omega_1,\Omega_2$, and observation of the
particle in $\Omega_1$ leads to instantaneous reduction of the wave function
irrespective to the distance between the domains.

Now, compare this state vector
with the two-particle excitation $\hat a(f_1)^+\hat a(f_2)^+|0\rangle$. Here
the field is also excited in domains $\Omega_1,\Omega_2$, but
$||f_{1,2}||^2\neq 1$, i.e. $\hat a(f_{1,2})^+|0\rangle$ cannot be considered
as one-particle states. Such states can be created from one-particle states
by splitting them. For photons, as a splitter physicists use semisilvered
mirrors.

All this is true for a macroscopic body $B$, which is also some excitation
of fields. Let $\hat B^+[\psi_i], i=1,2$, be the creation operator of the
body with the center of mass coordinates ${\bf X}_i$, such that the domains
where $\psi_i\neq 0$ do not intersect. Then the state vector
$|\psi\rangle=\hat B^+[\psi]|0\rangle$, $\psi=\psi_1+\psi_2$
($||\psi||^2=||\psi_1||^2+||\psi_2||^2$), describes the superposition of the
body states with different centers of mass ${\bf X}_1,{\bf X}_2$. In case of
measurement the "splitter" is nothing but the microscopic object
$o$. Of course, it cannot split a macroscopic object.

In this consideration the body consists of finite number of particles and
both states $|\psi_{1,2}\rangle$ belong to the same QFT Hilbert space. But
in the limit $N\rightarrow \infty$ they belong to unitary non-equivalent
Hilbert spaces. It allows to see the difference between mathematical
(formal) and physical approaches to the same phenomenon. We should admit
that for macroscopic bodies there are {\it weak superselection rules}. In
principle, there exists an operator preparing superposition
$|\psi_{1}\rangle + |\psi_{2}\rangle$ (a "splitter"). But even for a photon
it is a complex device. For macroscopic bodies the splitting is much more
difficult task (see below about "the Schroedinger cat"). In the case of the
process (1) the microscopic object $o$ cannot split the wave function of the
measuring apparatus. In practice it leads to SR for macroscopic objects.

{\bf 4.}
To demonstrate limitations of quantum mechanics, E.~Schroedinger proposed a
thought experiment [12] with a cat in a superposition of a dead and alive
states. In [13] the Schroedinger idea was realized, though without animals.
There was observed a superposition of two currents flowing clockwise and
anticlockwise. The number of electrons involved in the currents can be
estimated as $N_0\sim 10^{10}$. Question: does this experiment contradicts
to the Theorem? The latter is valid in the limit $N\rightarrow\infty$, so
one has to decide, is the number $N_0$ "macroscopically meaningful" or not.
The experiment [13] shows that it is not. A body with mass of a typical
macroscopic object $M\sim 10^{-3} kg$ consists of $\sim 10^{23}$ atoms.  A
corpuscle consisting of $\sim 10^{10}$ atoms has mass $\sim 10^{-16} kg$.
Experimenters never used apparatuses of such masses.
The experiment [13] deals with the border between micro- and macro-physics
where the superposition of the quantum states is still detectable. But this
experiment is in accord with the "weak" SR. The role of splitter there
played the microwave bath. It would be interesting to study dependence of
observed in [13] effect on the number $N_0$ of electrons in the currents.

In conclusion, we see that the problem of measurement is connected with quite
different aspects of quantum mechanics. Here interplay both historical
"prejudices" and difficulties of the problem by itself:
macroscopic bodies consist of finite, though enormous numbers of particles,
while for their description physicists use methods of QFT --- the theory of
systems with infinite numbers of degrees of freedom. One of the main
problems in the issue of measurement is decoherence, i.e. transition from
r.h.s. in (1) to (2). This is possible in case of existence of SR, which are
well understood in case of systems with infinite numbers of degrees of
freedom (a ferromagnetic). If we {\it define} a
macroscopic body as an object having all the properties of a $N$-particle
system in the limit $N\rightarrow\infty$, then
a formal proof of SR for bodies with different centers of mass
is possible. Evidently, such a definition is unavoidable in any formal proof
operating with the notion of "macroscopic object". For real macroscopic
bodies the problem of superselection is connected with the problem of
splitting of their wave functions.

The author is grateful to Prof. E.~Merzbacher for a stimulating remark.

\end{document}